\documentclass[12pt]{article}

\usepackage{graphicx}
\usepackage{a4}

\usepackage{fancybox}
\usepackage{epsfig}
\usepackage{color}

\usepackage{amsfonts}
\usepackage{mathrsfs}
\usepackage{amssymb}
\usepackage{amsmath}

\usepackage{dcolumn}
\usepackage{bm}

\def\pa{\partial}

\def\be{\beta}

\def\de{\delta}
\def\ep{\epsilon}

\def\th{\theta}

\def\ka{\kappa}

\def\Ga{\Gamma}

\def\Om{\Omega}

\def\D{{\cal D}}

\def\L{{\mathcal{L}}}

\newcommand{\ben}{\begin{equation}}
\newcommand{\een}{\end{equation}}
\newcommand{\bea}{\begin{eqnarray}}
\newcommand{\eea}{\end{eqnarray}}
\newcommand{\ba}{\begin{array}}
\newcommand{\ea}{\end{array}}
\newcommand{\bit}{\begin{itemize}}
\newcommand{\eit}{\end{itemize}}

\newcommand{\vs}[1]{\vspace{#1 mm}}

\newcommand{\dsl}{\pa \kern-0.5em /}

\begin{document}

\topmargin 0pt \oddsidemargin 0mm

\vspace{2mm}

\begin{center}

{\Large Towards the merger of Hawking radiating black holes}

\vs{10}

 {\large Huiquan Li \footnote{E-mail: lhq@ynao.ac.cn} and Jiancheng Wang}

\vspace{6mm}

{\em

Yunnan Observatories, Chinese Academy of Sciences, \\
650216 Kunming, China

Key Laboratory for the Structure and Evolution of Celestial Objects,
\\ Chinese Academy of Sciences, 650216 Kunming, China

Center for Astronomical Mega-Science, Chinese Academy of Sciences,
\\ 100012 Beijing, China}

\end{center}

\vs{9}

\begin{abstract}

We discuss the merger process of binary black holes with Hawking
radiation taken into account. Besides the redshifted radiation to
infinity, binary black holes can exchange radiation between
themselves, which is first redshifted and then blueshifted when it
propagates from one hole to the other. The exchange rate should be
large when the temperature-divergent horizons are penetrating each
other to form a single horizon with unique temperature. This will
cause non-negligible mass and angular momentum transfer between the
black holes during the merging process of the horizons. We further
argue in the large mass ratio limit that the light hole whose local
evaporation is enhanced by the competing redshift-blueshift effects
will probably evaporate or decay completely before reaching the the
horizon of the heavy one. We also discuss the possibility of testing
Hawking radiation and even exploring the information loss puzzle in
gravitational wave observations.

\end{abstract}

%\textit{Keywords:}

%\textit{PACS:}

\section{Introduction}
\label{sec:introduction}
%%%%%%%%%%%%%%%%%%%%%%%%%%%%%%%%%%%%%%%%%%%%%%%%%%%%%%%%%%%%%%%%%%%%%%%%%%%%

The gravitational science has been brought into a new era since the
first observation of gravitational waves (GWs) from binary black
holes by the LIGO/Virgo collaborations
\cite{Abbott:2016blz,TheLIGOScientific:2016wfe}. The successful
discovery opens a new window for the observational channels. It
provids a powerful tool to probe the properties of black holes.

The confirmation of a GW event needs the match of the signal with a
large amount of waveform templates. The physical parameters of the
original black holes and the final one are meanwhile inferred from
the matched waveform. The templates are evaluated in general
relativity with a combination of the analytical and numerical
methods. This is usually based on the classical mechanics of the
binary system, in which the black holes are usually taken as
point-like particles that only have masses and spins.

However, black holes are objects of finite size with an event
horizon that can radiate. The black hole mechanics takes close
analogy to the thermodynamical laws. This inspires the findings that
black holes have the thermal features, like entropy and temperature.
In the semiclassical theory, it was shown by Hawking
\cite{Hawking:1974sw} that black holes can radiate in the black body
spectrum from their event horizons. So black holes should evaporate
and evolve due to the Hawking radiation. As observed far from the
hole, the Hawking temperature is quite small and negligible for a
massive black hole so that it has a long lifetime
\cite{Page:1976df,Page:1976ki}.

It has been pointed out that the mechanical laws for isolated black
holes can be extended to the case of multiple black hole systems
\cite{Friedman:2001pf}. This is further verified in various
examinations in post-Newtonian (PN) and field theory approaches
\cite{Uryu:2010su,LeTiec:2011ab,Zimmerman:2016ajr,LeTiec:2017ebm}.
It implies that thermal features should still hold in multiple black
holes. The black holes should also evolve due to radiation. It is
easy to find that, except for the radiation escaping to infinity,
the black holes in the multiple system could exchange Hawking
radiation among themselves.

So, involving the quantum features of black holes, we should take
Hawking radiation into account in considering the dynamics and GW
emission of binary black holes, as suggested in
\cite{Giddings:2016plq}. Of course, the Hawking radiation from a
massive black hole is quite weak as observed just some distance away
from the horizon. The thermal particles emitted from the event
horizon will be mostly reflected back by the curvature of the
spacetime and only very few of them with high energies can escape
from the near-horizon region. But the radiation exchanged between
the black holes would be not necessarily weak, in particular when
the black holes get very close to each other. For a stationary black
hole, the local Hawking temperature increases as distance decreases
and diverges at exactly the horizon. So, when considering the merger
of black holes, we must pay additional attention on the crossing
process of the temperature-divergent horizons, during which the
exchange of thermal particles could be non-negligible.

This can also be understood in the ``thermal atmosphere" scenario
\cite{Thorne:1986iy}. Due to Hawking radiation, the space around a
black hole is not empty for local observers near the horizon in the
semiclassical theory. In the membrane paradigm, the thermal
particles emitted and trapped in the narrow layer just above the
horizon form a ``thermal atmosphere", which can be viewed as a
surrogate of the black hole. It recodes all the information of the
hole in its history, including the mass, angular momentum and
entropy. So the merging process of binary black holes can be viewed
as the mixing process of two thermal subsystems. There should exist
mass and angular-momentum transfer between the two atmospheres.

The exchange of Hawking radiation will cause the black hole
parameters to vary with time. In contrast to the classical theory
case, the masses and angular momenta of the black holes may not be
constant any more. In particular, the variation of the black hole
parameters should become quite prominent when the
temperature-divergent horizons are penetrating each other. The
time-varying parameters should affect the waveforms of GWs emitted
from binary black holes with different mass ratios. So we might
discriminate these features from the GW signals and test the Hawking
radiation effects.

However, the aim to tackle exact Hawking radiation from binary black
holes is almost impossible at the moment since the accurate
spacetime is unknown. In this work, we make some simple analysis by
assuming that Hawking radiation is a thermal feature associated with
the event horizons, which are though hard to be located in a
dynamical system. The radiation flux between black holes is
determined with analysis of the redshift factor on the spacetime. In
the dynamical spacetime of the binary system, the redshift factor
associated with the Killing vectors is usually not derivable. But,
we think that there could be two situations that discussion can be
simplified. The first is the case when the black holes are separated
by a large distance. The other case is the one of a binary system
with large mass ratio. In such a system, the various corrections to
the redshift factor can be dealt with in a controllable way, though
we only consider the probe approximation in this work.

The paper is organised as follows. In Section.\ \ref{sec:hr}, we
first introduce the Hawking radiation from a single, isolated black
hole. In Section.\ \ref{sec:merger}, we discuss generally the
exchange of Hawking radiation between two black holes when they get
closer and closer. At the final stage of the merger, this leads to
non-negligible mass and angular momentum transfer between the holes
and causes the black hole parameters to evolve. For a binary system
with unequal masses, the evaporation of the light hole will be
greatly enhanced in the near-horizon region of the heavy hole. In
Section.\ \ref{sec:enhancedevaporation}, we argue in the large mass
ratio case that the light hole will evaporate completely before
reaching the horizon of the latter. In the final section, we
summarise and make some extended discussions.

%%%%%%%%%%%%%%%%%%%%%%%%%%%%%%%%%%%%%%%%%%%%%%%%%%%%%%%%%%%%%%%%%%%%%%%%%%%%
\section{Hawking radiation of a single black hole}
\label{sec:hr}
%%%%%%%%%%%%%%%%%%%%%%%%%%%%%%%%%%%%%%%%%%%%%%%%%%%%%%%%%%%%%%%%%%%%%%%%%%%%

For a single black hole in empty vacuum, the temperature of Hawking
radiation observed at spatial infinity is proportional to the
horizon surface gravity: $T_{H}=\ka/2\pi$. The local temperature
observed by a stationary observer at any distance larger than the
horizon radius is (e.g., \cite{Wald:1984rg})
\begin{equation}\label{e:localtemp}
 T=\frac{T_H}{V},
\end{equation}
where the redshift factor $V=\sqrt{\xi_\mu\xi^\mu}$ with $\xi^\mu$
being the Killing vector of the spacetime. $\ka$ is constant on the
orbits of $\xi^\mu$: $\L_\xi\ka=0$. Specifically, it is constant on
a stationary horizon, obeying the zero-th law. The Killing vector is
null on the horizon. So the local temperature diverges on the
horizon.

For a Schwarzschild black hole with mass $\mu$, the surface gravity
$\ka=1/4\mu$ and the redshift factor is
\begin{equation}\label{e:Sredshift}
 V=\sqrt{1-\frac{2\mu}{r}}.
\end{equation}
The local temperature decreases as the distance increases due to the
gravitational redshift $V$. It is approximately equal to $T_H$ at a
distance of the scale $\sim \mu$ from the horizon. But it becomes
large at a distance $r=2\mu(1+\ep)$ ($0\leq\ep\ll1$) very close to
the horizon:
\begin{equation}\label{e:}
 T\simeq\frac{T_H}{\sqrt{\ep}}.
\end{equation}

For a free-fall detector (a probe) whose acceleration near the
horizon is
\begin{equation}\label{e:nhlocalacc}
 a\rightarrow\frac{\ka}{V},
\end{equation}
it will detect no thermal bath with the temperature
(\ref{e:localtemp}). In terms of the Unruh effect, it is reasonable
to infer that an infalling object moving with a local acceleration
less than the one (\ref{e:nhlocalacc}) will still see a thermal
bath, but with a lower temperature. For example, a massive object
that can not be viewed as a probe will not reach the local
acceleration (\ref{e:nhlocalacc}) near the horizon. So it will see
locally a thermal bath with a temperature less than the one given in
Eq.\ (\ref{e:localtemp}).

%%%%%%%%%%%%%%%%%%%%%%%%%%%%%%%%%%%%%%%%%%%%%%%%%%%%%%%%%%%%%%%%%%%%%%%%%%%%
\section{Towards the merger of radiating black holes}
\label{sec:merger}
%%%%%%%%%%%%%%%%%%%%%%%%%%%%%%%%%%%%%%%%%%%%%%%%%%%%%%%%%%%%%%%%%%%%%%%%%%%%

We now discuss generally the merger process of two Hawking radiating
black holes with when they move closer and closer. Here, we only
consider the non-spinning black holes for simplicity. It is
straightforward to extend the results to the spinning case.

Let us consider the binary black holes with raidii $r_M=2M$ and
$r_m=2m$, where $M$ and $m$ ($M\geq m$) are the masses of the black
holes. The black holes still evaporate to future infinity
$\mathcal{I}^+$. Meanwhile, they can also exchange Hawking radiation
between themselves. The radiation from the past horizon
$\mathcal{H}^-$ of one of the holes can go to the future horizon
$\mathcal{H}^+$ of the other hole. During this process, the
radiation is not always redshifted, but is first redshifted near the
starting hole and then blueshifted near the ending hole. There
should be a turning point for this redshift-blueshift transition on
each trajectory in between the two holes. The locations of the
turning points should change as the binary system evolves. The total
redshift effect on a light ray from one horizon to the other is a
competing result of the redshift effects around the two holes
(notice that the redshift factors may be modified in the binary
system) since the redshift factor of one of the holes serves as the
blueshift factor for the other hole.

As we shall discuss in detail below, the counter effects can make
the Hawking radiation exchanged between black holes more important
than the radiation to infinity. For the thermal radiation, we
determine the radiation flux by comparing the temperatures, which
are more conveniently measured at the turning points (or surface).
Since the redshift factors of the two black holes are truncated by
each other\footnote{For a single black hole, the redshift factor
varies from 1 to infinity as $r$ decreases from infinity to the
horizon radius. In the presence of a companion hole, it can only
vary from a value larger than 1 to infinity as $r$ decreases from
the turning point to the horizon radius.}, the local temperatures
observed at a turning point should be larger than the one observed
at infinity. They will be large when the separation of two holes
becomes very small.

The exchange of Hawking radiation between black holes will induce
the evolution of the black hole parameters, like the masses and
spins. The thermal particles radiated from one hole can carry some
portions of mass and angular momentum away to the other hole. This
causes the mass and angular momentum transfer between the black
holes, which results in the time-varying parameters according to the
first law of mechanics.

\subsection{Large separation analysis}
\label{sec:largesep}

Initially, the two black holes are separated by a large distance
(much greater than their horizon radii), inspiralling around each
other with slow velocities. So the black holes may be approximately
taken as two isolated holes separated from each other. Their
relative motion and tidal interactions can be neglected. In this
case, their redshift factors $V_M$ and $V_m$ simply take the same
form (\ref{e:Sredshift}) as in the isolated case. The radiation
following any trajectory from the hole $m$ propagating to the one
$M$ is first redshifted and then blueshifted beyond a turning point.

We consider the turning point on the trajectory that coincides with
the straight line crossing both black hole centres, on which the
exchange rate of Hawking radiation is the largest. The position of
the turning point can be estimated by the balancing condition of the
counter redshift effects
\begin{equation}\label{e:}
 V_M\simeq V_m.
\end{equation}
Set the separation of the two horizons to be $D$ and the distance of
the turning point to the horizon of the hole $m$ to be $d_T$. From
the condition, the turning position is determined to be at
\begin{equation}\label{e:}
 d_T\simeq \frac{m}{M+m}D.
\end{equation}
So the turning point is closer to the smaller hole. This is also the
position where the gravitational forces of the two holes are equal
so that a ``stationary" observer can reside at this point. As the
holes spiral inwards, the turning position changes accordingly.
Thus, the observer is not exactly stationary and is taken to be
quasi-stationary if the system evolves slowly at large separation.
Counting in the rotation of the binary system, this point should be
replaced by the first Lagrangian point between the two holes.

At the turning point, the redshift factors are truncated to be
\begin{equation}\label{e:counterRfactor}
 \frac{1}{V_M}\simeq\frac{1}{V_m}\simeq\sqrt{1+\frac{r_M+r_m}{D}}.
\end{equation}
For finite $D$, the temperatures are larger than the corresponding
ones observed at infinity. When $M=m$, the black holes have the same
temperature and the net radiation flux between the black holes is
zero, i.e., each hole radiates and absorbs the same amount of
thermal particles (neglect the radiation to infinity). When the
black holes have different masses: $M>m$, the difference of the
local temperatures $\triangle T=T^{(m)}-T^{(M)}>0$. This means that
there is a net flux of thermal particles from the small hole to the
large hole.

As the separation $D$ decreases, the distances $d_T$ and $D-d_T$ to
the horizons of the two black holes both decrease. So the redshift
factors $V_m$ and $V_M$ are both more deeply truncated. The local
temperatures and their difference at the turning point increase as
$D$ decreases. But, in the large separation limit, the temperatures
and their difference are very small for massive black holes.

\subsection{Formation of ``common envelopes"} % of black spheres}

Due to Hawking radiation, a black hole is expected to evaporate
completely in a finite time. The lifetime of a black hole is
calculated in \cite{Page:1976df,Page:1976ki} by summing up the
emission rates of all species of particles that exist in nature.
Thermal particles radiated from the horizon will be possibly
reflected back by the curvature of the black hole spacetime. This
gives rise to an absorption cross section for each kind of the
particles, which characterises a black sphere. The black body
radiation of a kind of particle is viewed to be emitted from the
surface of the black sphere. At high energies, the radius of the
black sphere for all kinds of particles is $\sqrt{27}\mu$ for a hole
with mass $\mu$.

In a binary system, either hole contains a ``black sphere" whose
shape may be modified. When the separation $r_M+r_m+D$ between the
black holes becomes smaller than the total radii of the two black
spheres, there should form a ``common envelope" filled with the
corresponding particles. This common envelope is like the one formed
in binary systems of ordinary stars when their Roche lobes are both
filled with stellar material. Within the black hole common envelope,
the particles radiated from one of the holes are possibly absorbed
by itself or by the companion hole. If the black holes have
different temperatures, the low-temperature (high-temperature) hole
will absorb more (less) particles than it radiates. But the
communication rate of particles in the common envelope is still low
if the separation $D$ is not small enough.

\subsection{The merger of the ``temperature-divergent" horizons}

With the distance decreasing, the above analysis based on isolated
black holes is not applicable. In this case, the dynamics of the
system is non-perturbative and is usually studied numerically. The
black holes may attain relativistic velocities and the tidal
interactions between them become strong. It is found in numerical
studies that delicate phenomena emerge in this process
\cite{Hamerly:2010cr,Emparan:2016ylg,Hussain:2017ihw,Emparan:2017vyp,
Pook-Kolb:2019iao}.

In this case, the radiation between the black holes will be
complicated, which is not achievable at the moment since the highly
dynamical spacetime is unknown. Here, we make only some simple
analysis by still approximately taking the black holes as ordinary
moving and radiating objects. During the merging process of the
horizons, the exchange of Hawking radiation should be very strong
because the local temperatures become extremely high approaching the
horizons (Note that the horizons are deformed and develop spikes
during this process
\cite{Hamerly:2010cr,Emparan:2016ylg,Hussain:2017ihw,Emparan:2017vyp}).
Then the communication rate of Hawking radiation particles between
the two holes will become infinitely large, if the semi-classical
theory is still trustable in this case. This process proceeds till
the merging horizons reach a thermally equilibrium horizon with
unique temperature.

Hence, the mass and angular momentum transfer between the holes will
be very strong if the divergence of the local temperatures can not
be avoided. It was shown that, in the presence of a companion
object, the black hole will be cooled by the tidal interactions
\cite{Gralla:2012dm,LeTiec:2017ebm}. In what follows, we shall argue
that the tidal interactions and relative motion between black holes
are unable to effectively attenuate the extremely high exchanging
rate of Hawking radiation.

\subsubsection{Tidal force}

In the binary system, the surface gravity of a black hole is in
opposite direction to the tidal force from its companion hole. So
the surface gravity or Hawking temperature may be lowered due to the
tidal interactions. But, as we argue below, this is not enough to
avoid the divergence of the temperatures and their difference on the
touching horizons.

First, the divergence of the temperatures on the horizons should not
be avoided by tidal interactions. The tidal force by the companion
black hole can not reduce the surface gravity on the horizon to zero
because zero temperature is not achievable according the third law
of thermodynamics. For a non-zero temperature, the local Hawking
temperature on the horizon is still divergent since the Killing
vector in the redshift factor is null on the horizon.

Second, the huge difference of the local temperatures between
merging black holes can not be narrowed down by the tidal forces.
The radius of curvature near a black hole with mass $\mu$ is
$\mathcal{R}\simeq (r/r_H)^{3/2}r_H$, where the radius of the
horizon $r_H=2\mu$. So, in the binary system, the tidal force on the
heavy hole by the light hole is stronger than the reverse one. It
has been examined in the numerical simulations for extreme mass
ratio systems \cite{Emparan:2016ylg}. The light black hole moving
near the large black hole horizon is just like that it moves in a
flat spacetime since the curvature near the heavy hole is quite
small, while the horizon of the heavy hole is strongly deformed by
the light black hole. This means that the heavy hole will be cooled
more than the light one. Thus, the tidal interactions will increase
the temperature difference rather than smooth it.

\subsubsection{Motion}
\label{sec:motion}

The relative motion includes two aspects: the acceleration and
velocity.

\textit{$\bullet$ Acceleration}

The acceleration of the detector is crucial for it to detect the
thermal particles near the horizon. Observers moving with different
accelerations will see thermal bath of different temperatures, as
stated in the previous section. A stationary observer hovering near
horizon will see a thermal bath with the local temperature
(\ref{e:localtemp}), while a free-fall probe-like observer whose
acceleration is given by (\ref{e:nhlocalacc}) will see nothing at
all when crossing the horizon. The situation for two black holes is
different.

When the two black holes have comparable masses, the smaller hole
$m$ can not be viewed as a probe of the larger hole $M$. So its
free-fall acceleration can not reach the one (\ref{e:nhlocalacc})
locally when its horizon starts to touch the one of the larger hole
(Moreover, at the moment, the center of the hole $m$ is $\sim r_m$
away from the horizon of the latter, which also greatly suppresses
its near-horizon acceleration). For this case, the free-fall small
hole $m$ will still ``see" a thermal bath, but with a temperature
somehow lower than (\ref{e:localtemp}). But this can not remove the
divergence of the temperature though it is lower.

When the mass ratio is large $M\gg m$, the smaller hole can be
viewed as a probe. When it freely collapses towards $M$, it will see
no particles when crossing the horizon of $M$. But a stationary
observer hovering over the horizon of $M$ (the larger hole is almost
static since $M\gg m$) can see that the smaller hole $m$ is
radiating particles. The acceleration of the observer, together with
$M$, moving towards $m$ is equal to the one of $m$ falling towards
$M$ and it is much less than the intrinsic near-horizon acceleration
(given by Eq.\ (\ref{e:nhlocalacc})) of a probe falling towards $m$.
So the observer can see the hole $m$ is radiating at almost the
temperature (\ref{e:localtemp}) locally.

\textit{$\bullet$ Velocity}

The relative velocity is also important for the observed radiation.
The temperature $T=T_H$ from Eq.\ (\ref{e:localtemp}) is the one of
a single black hole observed by a static observer at spatial
infinity. We now suppose that the black hole is moving towards the
observer or equivalently the observer is moving towards the black
hole with a constant velocity. Then the observed radiation is
boosted by a Doppler factor.

This situation also occurs between the inspiral black holes. At
their merging stage, the relative velocity between the black holes
can reach the relativistic regime. For example, the relative
velocity (divided by the speed of light) in the first observed GW
event \cite{Abbott:2016blz} is nearly $\be\sim 0.6$ at the end of
the merger. The relative velocity between the holes will provide an
extra redshift factor for the Hawking radiations exchanged between
themselves.

We may estimate the effect by recalling the case of the radiation
from a radiating and moving source in Minkowski spacetime, though
the spacetime in our case is highly curved. The Doppler factor for
the observed radiation by a static observer is:
\begin{equation}\label{e:}
 \D=\frac{1}{\Ga(1-\be\cos\th)},
\end{equation}
where $\be$ is the velocity of the radiating source relative to the
observer and $\Ga=1/\sqrt{1-\be^2}$ is the Lorentz factor. The angle
$\th$ is the one between the moving direction of the source and the
line of sight of the observer. The observer will see that the
temperature is shifted by $T\rightarrow \D T$. In our case, we can
choose one of the black hole as the observer and the other as the
moving and Hawking radiating source.

For head-on collision of the black holes, the angle is $\th=0$. The
Doppler factor is $\D=(1+\be)\Ga$. So the Hawking radiation from one
of the holes to the other is boosted in head-on collision.

For the circular rotation case, the angle is $\th=\pi/2$ and so the
Doppler factor is $\D=1/\Ga$. In this case, the Hawking radiation
between the holes is de-boosted. But this redshift effect is finite.
For binary black holes with relative velocity $\be=0.9$, the Doppler
factor is not quite different from the case with zero velocity. Even
for the ultra-high velocity $\be=0.99$, the Lorentz factor is
$\Ga\simeq7$. But this case seems not easy to happen because it
needs tremendous energy to power a massive black hole to reach such
a high velocity. So the circular motion is also unable to
effectively attenuate the strong Hawking radiation exchanged between
merging black holes.

%%%%%%%%%%%%%%%%%%%%%%%%%%%%%%%%%%%%%%%%%%%%%%%%%%%%%%%%%%%%%%%%%%%%%%%%%%%%
\section{The extreme mass ratio limit}
\label{sec:enhancedevaporation}
%%%%%%%%%%%%%%%%%%%%%%%%%%%%%%%%%%%%%%%%%%%%%%%%%%%%%%%%%%%%%%%%%%%%%%%%%%%%

As the Hawking radiation is exchanged, there should exist mass and
angular momentum transfer between the black holes, which causes the
black hole parameters to evolve. To access the details of the
evolution of the parameters, numerical calculations are needed for
the general case. But the discussion may be simplified in the case
for two non-spinning black holes in the extreme mass ratio limit, in
which we could take the probe approximation. This case is applicable
to the binary systems of a supermassive galactic black hole orbited
by a stellar black hole, which are targets of future LISA detectors
\cite{Audley:2017drz}.

For a large mass ratio, the background spacetime is dominated by the
one of the heavy hole and is perturbed by the light hole at the
order $\mathcal{O}(m^2)$ \cite{Detweiler:2000gt}. The light hole
looks like a particle of size $\sim m$ immersed in the background
spacetime and it only governs the spacetime that is close to it
\cite{Emparan:2016ylg}. In the probe limit, we assume that the
spacetime of the heavy hole is rigid and unaffected by the light
hole. We ignore the tidal interactions between the holes, which
actually increases the difference of the local temperatures as
argued in the previous section, and the horizon deformations
\cite{Hamerly:2010cr,Emparan:2016ylg,Hussain:2017ihw,Emparan:2017vyp}
when the horizons moves towards touching each other.

We shall mainly focus on the process that happens near the horizon
of the heavy hole. This region is within the innermost stable
circular obit (ISCO), beyond which the light hole will plunge into
the heavy hole. Its velocity should be radially dominated. In this
case, the relativistic motion can actually boost the exchange rate
of Hawking radiation. But, we here ignore this Doppler boosting
effect from the relative motion.

As analysed in Section.\ (\ref{sec:motion}), the free-fall hole $m$
will receive almost nothing in the near-horizon regions of the heavy
hole $M$. We only need to consider the radiation received by the
heavy hole from the light one. With the above approximations, the
turning point between the holes (we only consider one turning point,
through which the exchange rate is the largest) could be determined
by the counter redshift effects from the two Schwarzschild black
holes. The redshift factors take the same form as given in Eq.\
(\ref{e:counterRfactor}). With large mass ratio, the turning
position is at
\begin{equation}\label{e:}
 d_T\simeq \frac{r_m}{r_M}D.
\end{equation}
The local temperature of the light hole with the redshift factor at
the turning point is
\begin{equation}\label{e:lightlocaltemp}
 T^{(m)}\simeq T_H^{(m)}\sqrt{1+\frac{r_M}{D}}.
\end{equation}
The temperature $T_H^{(m)}$ is associated with the intrinsic surface
gravity of the light hole measured in the isolated case.

When $D\gtrsim r_M$, the local temperature is nearly equal to
$T_H^{(m)}$. When $D\ll r_M$, the local temperature can be high:
$T^{(m)}\simeq\sqrt{r_M/D}T_H^{(m)}$. For example, the mass ratio of
a galactic black hole to a stellar one can reach $M/m=10^7$. The
local temperature with a separation $D=10r_m$ is $10^3$ times larger
than $T_H^{(m)}$. This means that its evaporation rate is almost
$10^{12}$ times larger.

As the light hole moves closer to the heavy hole, the evaporation
becomes more strongly enhanced since the redshift factor is more
deeply truncated. It will lose mass more quickly, which makes it
even hotter. The local temperature (\ref{e:lightlocaltemp}) diverges
at the horizon of the large hole ($D\rightarrow0$), consistent with
the analysis in the previous section. Thus, the light hole will
finally evaporate completely before reaching the horizon of the
heavy hole since the evaporation rate diverges there.

This can be clearly seen with some simple calculations. Near the
horizon of the heavy hole, the light hole moves at relativistic
speed and so $D\simeq -t+\textrm{const}$ with the local time
interval $d\tau=V_M(d_T)dt$. Then the evolving equation of the mass
of the light black hole towards the merger with the heavy hole can
be expressed as
\begin{equation}\label{e:evaporationeq}
 \frac{dm}{dD}\propto \de\Om r_m^2 {T_H^{(m)}}^4
\left(\frac{r_M}{D}\right)^{\frac{3}{2}},
\end{equation}
where $\de\Om$ is the solid angle of the Hawking radiation from the
the light hole that is effectively enhanced. In the large mass ratio
limit, the horizon of the heavy hole looks likes a plane compared
with the light hole. So we may assume that the effective solid angle
$\de\Om$ would be almost the same for the light hole with different
mass $m$.

We can set the mass $m$ at some initial point $D=D_0$ as $m_0$. The
mass becomes $m_1$ at point $D=D_1$ ($D_0>D_1$). Then we have from
the integration of the above equation:
\begin{equation}\label{e:}
 m_1=\left[m_0^3-K\left(\frac{1}{\sqrt{D_1}}-
\frac{1}{\sqrt{D}_0}\right)\right]^{\frac{1}{3}},
\end{equation}
where $K$ is a positive constant. So the mass decreases as the
separation $D$ decreases. The light hole will evaporate completely
with $m_1=0$ at
\begin{equation}\label{e:}
 D_1=\left(\frac{1}{\sqrt{D_0}}+\frac{m_0^3}{K}\right)^{-2},
\end{equation}
before the horizon of the heavy hole is reached (i.e., $D_1>0$). The
evaporated particles are mainly absorbed by the large hole. Note
that this locally enhanced evaporation proceeds near the horizon of
the heavy hole and it may still remain weak as observed by an
observer at infinity due to the huge gravitational redshift. But the
observer can sense the occurrence of this mass transfer process by
observing the dynamics and GW emission of the binary system.

The result is quite different from that in classical theory, but is
similar to the speculations in the Rindler/Tachyon correlation
proposed in our previous works \cite{Li:2011ypa,Li:2014jfd} (and its
extension to the dS space \cite{Li:2015qmf}). The mysterious
similarity between the action of a probe particle in Rindler space
(the geometry near the horizon of a non-extreme black hole) and the
tachyon effective action may suggest that the collapsing process of
matter towards a black hole horizon is a tachyon condensation
process, with the Hawking temperature equal to the Hagedorn
temperature in tachyon field theory. The self-energy of the particle
will leak to the background spacetime and will be absorbed by the
black hole as it moves toward the horizon. In terms of the tachyon
field theory, the particle will eventually decay completely into
massless and massive closed strings (or gravitons in pure gravity)
before reaching the horizon. Our discussion here thus extends this
suggestion to the case of particle-like black holes: a light hole
moving towards a massive hole will evaporate or decay completely
before reaching the horizon. The difference is that here the
``particle" is radiating and has temperature.

It is unknown whether the result holds beyond the large mass ratio
limit. We guess so if the divergence of the temperature (difference)
can not be avoided. But this awaits for critical examinations in
full-theory numerical simulations.

%%%%%%%%%%%%%%%%%%%%%%%%%%%%%%%%%%%%%%%%%%%%%%%%%%%%%%%%%%%%%%%%%%%%%%%%%%%%
\section{Conclusions and discussions}
\label{sec:conclusions}
%%%%%%%%%%%%%%%%%%%%%%%%%%%%%%%%%%%%%%%%%%%%%%%%%%%%%%%%%%%%%%%%%%%%%%%%%%%%

We consider the merger process of binary black holes when Hawking
radiation is involved. The black holes in a binary system should
radiate to each other, leading to mass and angular momentum transfer
between them. This quantum effect could be prominent in two
situations. One is the case that at least one of the black holes has
a high temperature (like the low-mass PBHs \cite{Page:1976wx}). The
other is at the final moments of the merging process of general
black holes.

As learnt from common sense, the exchange of Hawking radiation
particles, which include all kinds of particle in nature, will exert
a force or a torque on each of the black hole. This torque tends to
slow down the rotation of the binary system so that the black holes
can more effectively get closer. The force is the Casimir style
since the creation of thermal particles from the event horizons
resembles the one on a moving mirror via the dynamical Casimir
effect \cite{DeWitt:1975ys} (they may be essentially the same
\cite{Nugaev:1979fj}). The Casimir-Polder force between black holes
due to the exchange of gravitons have been discussed in
\cite{Ford:2015wls,Hu:2016lev}. It is also expected that the Casimir
force between black holes due to exchange of photons could exist
since the black hole horizons can be viewed as conductors with
surface resistivity $4\pi=377$ ohm
\cite{1978MNRAS.185..833Z,Damour:1978cg,Thorne:1986iy}.

The exchange rate of radiation will get extremely large when the
temperature-divergent horizons are penetrating each other, even for
the binary black holes with the same temperature. We argue that the
divergence of the Hawking temperature on the horizons is unable to
be attenuated by tidal interactions and relative motion between the
black holes. For binary black holes with different temperatures, the
difference of the local temperatures will diverge once their
horizons touch, which implies that the horizons will instantly
evolve into a thermal equilibrium horizon due to the huge difference
of the temperatures. The divergence may be hidden from far observers
by a common outer horizon formed before the black hole horizons
touch each other \cite{Anninos:1994pa,Pook-Kolb:2019iao} in a highly
dynamical system.

We find that it is interesting to discuss the merger of binary black
holes with large mass ratio. In the binary system, the evaporation
of the light hole is greatly enhanced in the background spacetime of
the heavy hole. When the light hole moves in the large spacetime of
the heavy hole, it will probably evaporate completely before
reaching the horizon of the latter. This is surprisingly consistent
with our speculations of the Rindler/Tachyon equivalent
descriptions.

The observation of the GWs provides a unique method to probe the
nature of black holes and to test gravities beyond GR (e.g., see the
reviews \cite{Berti:2018cxi,Barack:2018yly}). The exchange of
Hawking radiation will cause the parameters of the black holes to
evolve, which will make the GW waveforms different from those
predicted in classical theory of gravity\footnote{For example, at
the final stage of the merger of two Schwarzschild black holes, the
stain of the GW involving Hawking radiation effect should be less
than the one predicted in the classical theory since the discrepancy
of their masses becomes larger as the radiation exchange proceeds.}.
It is hopeful that we may test it in future precise GW observations.
However, the effect is prominent only at final stages of the merging
process of the horizons, which lasts for a short timescale.
Moreover, the inferred masses (and spins) of the black holes from
the signal waveforms are not well determined (at a degree of $90 \%$
for the first observed event \cite{TheLIGOScientific:2016wfe}). All
these challenge the test of Hawking radiation in GW observations.

The situation may be improved in binary systems with large mass
ratios, which are potential arena for testing fundamental physics
\cite{Berti:2019xgr,Berry:2019wgg}. We have shown that the light
hole would significantly lose mass when approaching the near-horizon
region of the heavy hole. We also show that the light hole whose
evaporation is greatly enhanced will completely evaporate into the
heavy hole finally before reaching the horizon of the latter. So it
is possible to explore the information loss paradox in the GW
observations from such binary systems since the the evolution of the
light hole during its whole life is recoded in the GW signals.
People have long been puzzled by the information loss problem. When
matter collapse into a black hole, its information is lost forever,
which violates the laws of quantum mechanics. Many theories have
been proposed to solve the problem
\cite{Preskill:1992tc,Polchinski:2016hrw,Unruh:2017uaw,Marolf:2017jkr}.
Some of them are related to the final stage of the evaporation at
the Planck scale, at which a remnant is left or the hole bursts.
This could be observable via GWs if the light hole indeed
``dissolves" into a large hole.

Before making the precise judgement, we need first to develop
detailed theories on the merger process of radiating black holes and
produce the corrected waveforms of GWs with numerical relativity in
the strong gravity regime.

\section*{Acknowledgements\markboth{Acknowledgements}{Acknowledgements}}

This work is supported by the Yunnan Natural Science Foundation
2017FB005 and 2014FB188.

%\newpage
\bibliographystyle{JHEP}
\bibliography{b}

\end{document}